\documentclass[aps,prb,twocolumn,secnumarabic,amsmath,amssymb,a4]{revtex4}

\usepackage{graphicx}% Include figure files
\usepackage{dcolumn}% Align table columns on decimal point
\usepackage{bm}% bold math
\usepackage{fancyhdr}

\begin{document}
\fancyfoot[RO, LE] {Contemporary Physics, to appear}
\baselineskip 1em 
% \preprint{APS/123-QED}

\title{Undulatory Locomotion}% Force line breaks with \\

\author{Netta Cohen}
 \email{n.cohen@leeds.ac.uk}
\affiliation{
School of Computing and Institute of Systems and Membrane Biology, 
University of Leeds, Leeds LS2 9JT, United Kingdom
}%

\author{Jordan H. Boyle}
 \email{jboyle@comp.leeds.ac.uk}
\affiliation{%
School of Computing, University of Leeds LS2 9JT, Leeds, United Kingdom
}%

\date{18 August 2009}% It is always \today, today,

\begin{abstract}
Undulatory locomotion is a means of self-propulsion that relies on the
generation and propagation of waves along a body. As a mode of locomotion it is
primitive and relatively simple, yet can be remarkably robust. No wonder then,
that it is so prevalent across a range of biological scales from motile
bacteria to gigantic prehistoric snakes. Key to understanding undulatory
locomotion is the body's interplay with the physical environment, which the
swimmer or crawler will exploit to generate propulsion, and in some cases, even
to generate the underlying undulations. This review focuses by and large on
undulators in the low Reynolds numbers regime, where the physics of the
environment can be much more tractable.  We review some key concepts and
theoretical advances, as well as simulation tools and results applied to
selected examples of biological swimmers. In particular, we extend the
discussion to some simple cases of locomotion in non-Newtonian media as well as
to small animals, in which the nervous system, motor control, body properties
and the environment must all be considered to understand how undulations are
generated and modulated. To conclude, we review recent progress in microrobotic
undulators that may one day become commonplace in applications ranging from
toxic waste disposal to minimally invasive surgery.
\end{abstract}

\pacs{Valid PACS appear here}% PACS, the Physics and Astronomy
\maketitle

\section{The universality of undulatory locomotion}

One of the crucial existential requirements of microbial and animal life
is movement. Whether for foraging, for mating or to avoid danger, most
non-plant life forms have evolved mechanisms of locomotion, and the
solutions are diverse. Microorganisms are too small to have many
components in their motor system. Larger terrestrial animals may resort to
limbs that support the weight of the animal
and thus deal with gravitational forces. Other animals have adopted a
variety of forms that keep them close to the ground or even underground,
minimizing the effects of gravity. One example is the sandfish (small lizards
with smooth scales), whose undulations effectively fluidize the surrounding sand 
and allow it to swim through the granular flow \cite{baumgartner08}. 
Marine, subterrainean
and flying species have to contend with very different physical
environments.  And yet, all of locomotion can probably be divided into
a small number of fundamental classes. Of these, undulatory
locomotion that relies on wave propagation along the body, is remarkably
widespread.

No wonder. To undulate, a system does not require limbs, only a body. Even
single celled microorganisms use elastic filaments that extend from the cell
membrane, or modulate the body shape, to undulate.  
Paramecia, like many single celled eukaryotic organisms (protozoa), 
are covered with cilia that are used much like little oars to synchronously 
row through water.
Motile bacteria use one or more long helical
filaments, called flagella, that extend from the cell membrane and act as a
propeller. These flagellated bacteria swim by rotating all their helical
flagella one way, while the cell body rotates in the opposite direction.  {\it
E.\ coli}, for example uses such a mechanism to swim at 35 diameters per second
\cite{berg00,berg04}.  Some motile unflagellated bacteria resort instead to
the propagation of low amplitude waves on the surface of their bodies
\cite{stone96,ehlers96}. 

In the animal kingdom, limbs might come in handy for some purposes, but when
they are more of an impediment, swimmers can choose to 
press them against the body or use them to complement the undulations. Crawlers
may need to burrow into narrow nooks. And many species may exploit the
flexibility and adaptability of crawling like gaits to allow them to move
through a variety of different physical environments (such as eels, that not
only swim but are also capable of burrowing through sand or mud \cite{breder26}).  
Thus, it is not surprising to find
undulations as the primary means of locomotion in a variety of larvae, worms,
lizards, snakes, fish, and even some mammals ranging in size from 
tiny microbes to the monster (13m long) prehistoric titanoboa snakes \cite{head09}.

Animals use a variety of different forms of undulation to locomote. These can
crudely be classified into horizontal or vertical, direct (in the direction of
motion) or retrograde (opposite the direction of motion) and longitudinal or
transverse. Fish usually flap their bodies from side to side,
whereas whales and other marine mammals undulate up and down.  Typically, to
move in a given direction, retrograde waves are required, that propagate
opposite to the direction of motion (i.e., from head to tail to achieve forward
motion). In other words, if the body wave travels backward, the environment
applies forward forces.  But Biology is full of exceptions.  Some annelids,
including the polychaete worm {\it Nereis virens} \cite{laspina07}, as well as
several protozoa, such as the motile alga {\it Ochromonas
malhamensis} \cite{jahn64} exhibit the remarkable property that the organism
moves in the {\em same} direction as the wave propagation. To understand how
such a counter-intuitive mode of locomotion can be achieved, one must first
notice a key similarity between the body of {\it Nereis} and the flagella of
{\it Ochromonas}. In both cases, the body/flagellum is lined with `bristles'
which jut out at right angles to the long axis. These protrusions act much like
oars, generating significant drag forces when the body is moved forwards or
backwards, but having minimal effect when it is moved sideways. In contrast,
conventional (retrodgrade) undulations rely on stronger resistance to 
sideways motion.

If there is a unifying principle to all of the above, it is that undulating
motion is typically constrained by frictional or drag forces of the
environment, rather than by gravitational forces. For instance, most swimmers
can float in water using buoyancy, but self-propulsion becomes much more energy
intensive than in air or on the ground. This is all the more severe in
turbulent waters or when swimming upstream in a river. Fish use a range of
individual and group strategies to exploit the hydrodynamics and minimize
energy expenditure.  
Many fish, for example, can minimize power expenditure (i.e., muscle work) by
recycling energy from vortices in turbulent flows. In fact, by swimming in
schools, trout kinematics are not so different from that of passive hydrofoils
(mimicking flag waving motion) \cite{muller03,liao03,liao07}.
Terrestrial crawlers (and swimmers \cite{baumgartner08,maladen09}) 
keep close to the ground,
so they are much more constrained by the terrain and its associated
frictional forces than by gravity. But this is perhaps most obvious in the so
called low Reynolds number regime, where inertial forces become
altogether negligible. In all of the above, the immediate implication is that
undulatory locomotion arises from the interaction between the dynamics of the
body and the physics of the environment and hence places strong constraints on
the shape of the body. 

\section{Theory}

In this paper, we will try to restrict ourselves to some of the simplest
cases of undulatory locomotion operating in the low Reynolds number regime 
and where two dimensional treatments suffice;
body shapes are long and slender, and the environment is relatively
simple. We begin with an introduction to some of the theoretical foundations 
in fluid mechanics and their applications to self-propulsion in general and to
undulatory locomotion in particular.

\subsection{What is the Reynolds number?}
\label{sec:Re}

Fluid mechanics has long been of interest to physicists.
Already Isaac Newton postulated how fluids of different consistencies
respond to forces.  Perhaps, when Isaac Newton took a break from pondering
the motion of falling apples, he was holding a spoon over his cup of afternoon
tea, and dragging it along the surface. He would have noticed that the top
layer of the fluid was dragged along. How much more difficult would this
be if he had done the same across a jug of honey? Newton postulated that
the force required to keep a flat spoon moving at constant speed $v$ would
follow 
\begin{equation} 
F \propto A \left.\frac{dv}{dy}\right|_{\text{moving spoon}}
\label{eq:viscosity}
\end{equation} 
where $A$ is the contact area of the spoon, and $v(y)$ is
the speed profile of different slices of fluid as one moves a distance $y$
away from the spoon.  The proportionality constant is called the
viscosity of the fluid (or sometimes the dynamic or Newtonian viscosity)
$\mu$.  The gradient of the velocity profile $dv/dy$ reflects the strength
of the dragging and indicates how nearby slices of fluid are
differentially dragged
\footnote{In addition to being linear, Eq.\ (\ref{eq:viscosity}) stipulates
that fluid that is infinitessimally close to the spoon will travel at the
same speed as the spoon. This is often called the no-slip boundary condition 
(see also Sec. \ref {sec:NavierStokes}).}.
In fact, not all fluids obey the linearity of Eq.\ (\ref{eq:viscosity}), but
those that do are called Newtonian fluids.  Examples include air, water
and indeed honey with respective viscosities of
$O(10^{-5})$Pa$\cdot$s, $O(10^{-3})$Pa$\cdot$s and $O(10^{4})$Pa$\cdot$s.

Now if we change the speed with which we are moving the spoon (or stop
it altogether), we must also consider the inertia of the fluid. Inertial
effects will dominate over viscous forces when the spoon is sufficiently
fast, or alternatively, when the fluid has sufficiently low viscosity. In
this case, the force applied by the spoon on the fluid can result in
nonlinear convective and turbulent flows, which will then feed back and
influence the motion. Let us try to determine when such inertial effects
are important.

To do so, consider the motion of an object through a Newtonian fluid.  Suppose
the velocity $v$ of the fluid drops off linearly away from the object. The
viscous forces in the fluid around the object are given by $\mu A dv/dy$, which
will then scale as $O(\mu \ell^2 v /\ell) = O(\mu \ell v)$, where $\ell$ is a
characteristic size of the object. The inertial forces (due to the fluid's
momentum in the same region $m dv/dt$) should scale as $O(\rho \ell^2 v^2)$
where $\rho$ is the density of the fluid.  The ratio of these two expressions
is then characterized by a single dimensionless scaling parameter 
\begin{equation} 
\text{Re}= \frac{\rho \ell^2 v^2}{\mu \ell v}
         =\frac{\rho \ell v}{\mu}\;, 
\label{eq:Re} 
\end{equation} 
where Re is the conventional shorthand for the `Reynolds number'. 
When $\text{Re}\gg 1$ inertial forces
dominate.  By contrast, if $\text{Re}\lessapprox 1$ the viscous forces
dominate the flow and the fluid largely responds to external forces in a
passive manner.  To give ballpark figures, a person swimming in water
might experience a Reynolds number of $O(10^4)$.  If we tried to swim
through honey, we might feel a Re around $O(10^{-3})$ and bacteria swimming
in water may feel Reynolds numbers as low as $O(10^{-5}$)!

The Reynolds number can also be obtained from the governing
equation in fluid dynamics, the so-called Navier-Stokes equation 
(given here in simpler form for incompressible fluids, i.e., 
$\nabla\cdot v=0$) 
\begin{equation}
-\nabla p + \mu \nabla^2 v = 
\rho \frac{\partial v}{\partial t} + \rho (v\cdot \nabla) v\;.
\label{eq:NavierStokes}
\end{equation}
Here, the left hand side describes pressure and viscous terms, and the right
hand side describes inertial terms, which vanish at low Re \footnote{Strictly
speaking the Reynolds number is defined as the ratio of the viscous to the
convective (nonlinear) term, but in the low Re regime, both inertial terms can
be neglected \cite{purcell77}.}.  In fact, it is easy to see here, that for an
object with a characteristic length $\ell$, we can recover the Reynolds number
as the ratio of the inertial term $\rho (v\cdot \nabla) v$ to the viscous drag
term $\mu \nabla^2 v$.  In most of what follows, we will need only the low Re
reduction of the Navier-Stokes equation 
\begin{equation}
\mu \nabla^2 v = \nabla p\;.
\label{eq:lowRe}
\end{equation}

\subsection{Self-propulsion in low Re environments and
the scallop theorem}
\label{sec:scallop}

Most of our intuition comes from our day to day experiences of the high Re
world in which inertia must be overcome. When you start off in your car, a torque 
is applied to the wheels, which in turn apply a backwards directed force to the
surface of the road. The reactive (forwards directed) frictional force on the
car gives rise to acceleration. At some speed this propulsive force is
counter-balanced by wind resistance and the velocity settles.  The elimination
of inertia at low Re means that {\em any} nonzero resultant force acting on an
object will give rise to infinite acceleration. Thus, somewhat counterintuitively, 
the total net force and torque acting on an object moving in a 
low Re environment will at all times be zero. As an example, suppose a small 
swimmer is
moving at constant velocity, and then stops swimming. The above condition will
result in immediate deceleration and the swimmer will stop. Swimming is indeed
hard work at low Reynolds numbers.  Curiously, low Reynolds number physics is
remarkably reminiscent of the Aristotelian view of physics, in which
objects will remain stationary in the absence of external forces. Aristotle's
mechanics has long been dismissed as fundamentally flawed and superceded by
Newtonian mechanics, so it is reassuring to see that this theory too has 
found its natural place.

Let us now consider a small swimmer (and hence at low Re). 
To have any chance of moving, it must be able to change its shape.
The sum of all internal forces
must clearly be zero (for the same reason you cannot lift yourself up by your
boot straps).  The change of shape will result in some motion of parts of the
body in a global coordinate frame, which will then elicit reactive drag forces.
But since these reactive forces {\em must} sum to zero, the organism as a
whole will move in such a way that this is the case. This condition is
sufficient to uniquely determine the motion of the whole organism given a known
time series of body configurations and known environmental properties. 

In fact the low Re physics imposes constraints on the possible shape changes
that will result in progress. This was realized by Ludwig \cite{ludwig30} and
then by Purcell \cite{purcell77} who nicely formulated it as {\it the scallop
theorem}. 
Consider a scallop that opens and closes its shell in water to move.
At sufficiently high Re, the slow opening and rapid shutting of the shell
pushes water out and propels the scallop in the opposite direction.  At low
Reynolds number, the flow of water into and out of the scallop over one cycle
would be the same, regardless of the speed.  Scallops would make no net
progress at low Re. 

The derivation of this theorem is straightforward and instructive.  We
begin with the Navier-Stokes equation at low Reynolds number [Eq.\
(\ref{eq:lowRe})].  Note that this equation is time independent. This
means that speed makes no difference to the motion. Only the sequence of
configurations of the body determines the motion. But if that sequence is
time reversible, such that moving forward or backward in time involves the
same sequence of shapes, then no overall progress can result. 

How then is propulsion achieved at low Re? It is reasonable to assume that
successful propulsion always relies on a repetitive ``stride'' or cyclical
motion,
but can we move beyond the no-go scallop theorem towards a unifying theory of
shape changes that sustain self-propulsion at low Re?

\subsubsection{A generalized scallop theorem}

Consider some cyclic motion of a low Reynolds number swimmer in a Newtonian
fluid. Following Shapere and Wilczek \cite{shapere89} we could construct an
alphabet of all possible shapes $\{S_0\}$ of the swimmer, all at the same
location and orientation (the origin). Then the motion of a deformable body
undergoing an ordered sequence of body-shape changes $S_0(t)$ will be described
by a rotation and displacement of the body to its appropriate orientation and
location $S(t)$ via 
$$S(t) = \mathcal{R} (t)S_0(t)\;,$$
where $\mathcal{R}$ combines both rotation and displacement operations.  Now to
follow shape changes in continuous time we would like to follow infinitessimal
shape changes, and hence it is convenient to introduce an exponential form (so
infinitessimal generators will form a Lie algebra). In particular, defining 
$$ 
\frac{d\mathcal{R}}{dt} = 
\mathcal{R} \left( \mathcal{R}^{-1} \frac{d\mathcal{R}}{dt} \right) \equiv 
\mathcal{R}A 
$$
allows us to write 
\begin{equation} 
\mathcal{R}(t) = \bar P \exp\left({\int_0^t A(t')dt'}\right)\;, 
\label{eq:Lie} 
\end{equation} where $\bar P$ represents ordering
terms in the expansion of the exponent with later terms to the right
\footnote{This is exactly analogous to the solution of the time-dependent
Schr\"odinger equation in terms of the time ordered exponential.}.  Although
$A(t)$ is in general time dependent, the linearity of the Navier-Stokes
equation at low Re means that speed does not matter.  Indeed, as we have
already argued, in the absence of inertial effects, the sequence of shape
changes completely specified the rotation and displacement of a body.  Thus,
$A(t)$ may be recast in an effectively time-independent geometric manner
\cite{shapere89}.  Define a vector over shape space, also denoted $A$, whose
projection onto the ``direction'' (in shape space) $dS_0/dt$ is just $A(t)$,
via 
$$ A(t)\equiv A_{\dot S_0} [S_0(t)]\;.  
$$ 
Thus, at low Re, an integral over time can be recast as an integral over 
shapes and Eq.\ (\ref{eq:Lie}) becomes
$$
\mathcal{R}(t) = \bar P \exp\left({\int_{S_0(0)}^{S_0(t)} A[S_0]\cdot
dS_0}\right)\;. 
$$
For a cyclic stroke, the net rotation and displacement are then
$$
\mathcal{R}(t) = \bar P \exp\left({\oint A[S_0]\cdot dS_0}\right)\;. 
$$
Thus, the net
progress our swimmer makes in each cycle is proportional to the area
circumscribed in shape space. For a simple back-and-forth motion such as the
scallop's, no area is ``cut out'' and so no overall progress can be made. This
geometric formulation therefore generalizes Purcell's scallop theorem by
relating the progress of a low Re swimmer to a geometric phase (akin to a
classical Berry's phase). Of course, to determine how much progress is actually
made still requires solving the fluid dynamic equations to obtain the shape
space vector $A[S_0]$. 
This generalized formulation of the scallop theorem is
due to Shapere and Wilczek \cite{shapere89}.

\subsubsection{Spherical symmetry breaking for propulsion} 

While violating the time reversibility condition of the scallop theorem is
necessary for successful propulsion at low Re, there is one other asymmetry
that needs mentioning. Regardless of scale or Reynolds number, successful
propulsion requires an asymmetry or anisotropy in the environmental resistance
to the motion of the body.  Consider a compact shape that undergoes a
time-asymmetric cycle of shape changes, but remains at all times spherically
symmetrical.  It is trivial to see that such a shape would go nowhere.  Thus,
some form of spherical symmetry breaking is needed to achieve locomotion. 

Indeed, nearly all life forms, from bacteria to mammals have a distinct 
body axis or polarity, which dictates the direction of motion. When an
elongated body is moving forwards (parallel to its long axis) at low Re, 
it will displace less fluid per unit time than when moving sideways 
(normal to its long axis). Thus, it will encounter a smaller resistance 
from the fluid. The ability of a deformable body to modulate the level of 
resistance it encounters in different directions allows, in principle, for 
motion to be possible. This point will be revisited in Sec.\ 
\ref{sec:long_slender}.

\begin{figure*}[htb]
\begin{center}
\includegraphics[width = 0.8\textwidth]{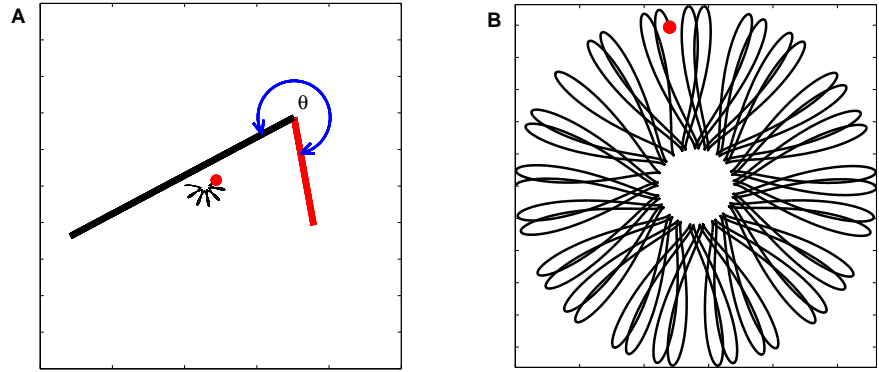}
\caption{ 
A. A variation on Purcell's scallop in which the two arms have different
lengths and are allowed to rotate freely around (like a wheel). Arms are
modeled as slender cylindrical bodies. The medium is modeled as Newtonian ($K$
= 2). The path of the scallop's ``center of mass'' (CoM, red dot) is shown in
black. After a few rotations, the CoM appears to undergo both translation and
rotation. However, the CoM never leaves a circular area of small radius, as can
be seen by following the trajectory over sufficiently many cycles (B). Each
`petal' corresponds to one cycle of rotation through $2 \pi$.}
\label{fig:scallop}
\end{center}
\end{figure*}

To demonstrate the minimal conditions for swimming, we give two very simple
examples: one of a generalized scallop that can rotate, but cannot swim (or
propel itself in one direction), and another of a minimal swimmer.  Consider a
cyclical version of Purcell's scallop, where
instead of opening and shutting, the scallop's opening angle keeps increasing
(at $2\pi$ per cycle). Shapere and Wilczek \cite{shapere89} already noted that
the Scallop theorem assumes that the scallop cannot turn through more than
$2\pi$ on its hinge). However, if we allow the scallop to do so, and assuming
the two arms are even, the scallop will still get nowhere.  To break time
reversal symmetry, one arm could be longer than the other, and hence subject to
a greater resistive drag force, as shown in Fig.\ \ref{fig:scallop}A. In this
case, the scallop can rotate, but will not achieve any lasting translation
(Fig.\ \ref{fig:scallop}B).

\begin{figure}[htb]
\begin{center}
\includegraphics[width = 0.8\columnwidth]{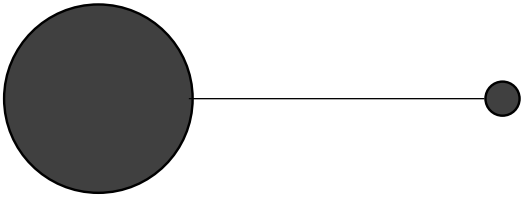}
\caption{Schematic of a self-propelled push-me-pull-you device, adapted from
Ref.\ [\protect{\onlinecite{avron05}}].  
To move, the device follows a cyclical sequence of steps that
include: (i) shortening the connecting rod between the spheres; (ii) inflating
and deflating the two spheres so their volumes are switched; (iii) lengthening
the rod to its original length; and (iv) deflating and inflating the two
spheres to their original sizes. For the initial conditions depicted here, this
sequence will lead to motion to the left.}
\label{fig:pushmepullyou}
\end{center}
\end{figure}

Consider, by contrast, the following push-me-pull-you example due to Avron 
{\it et al.} \cite{avron05} in Fig.\ \ref{fig:pushmepullyou}.
Here we have two perfectly spherical shapes that are experiencing isotropic drag
at all times. However, by allowing the spheres to inflate and deflate, the
relative drag forces they experience change in time, allowing the motion of the
spheres to be asymmetrical as well. 
In either case, applying Shapere's and Wilczek's formalism above should 
demonstrate that an area is indeed carved out in shape space, leading,
in the asymmetric scallop case, to an overall rotation and, in the 
push-me-pull-you case, to an overall displacement. (Note that the 
push-me-pull-you example involves two degrees of freedom, but some restricted
forms of propulsion can also be achieved with only a single degree of freedom.)

\subsection{Incompressible fluid dynamics at low Re}
\label{sec:NavierStokes}

To solve the equations of motion of a body moving through a fluid is a daunting
task which, in general, requires a solution to the Navier-Stokes
equations.  When convective and turbulent forces dominate, this is indeed
arduous , but even at low Reynolds numbers the problem is rarely
solved analytically. 
Below, we introduce the basic equations for relatively simple but limited
classes of problems. 
 We begin with the condition for incompressible fluids 
$\nabla\cdot v = 0$.  Taking the curl of Eq.\ (\ref{eq:lowRe}) yields
\begin{equation}
\nabla^2(\nabla\times v) = 0 \;.
\label{eq:incompressible1}%\\
\end{equation}
To solve the equation for swimmers in a fluid, we need only add boundary
conditions, satisfied by the fluid at the boundary of the swimmer's body. This
is typically the no-slip condition
\begin{equation}
v|_S=\frac{\delta S}{\delta t}\;,
\label{eq:noslip}
\end{equation}
which requires that the fluid is perfectly dragged along at the boundary $S$.

We now consider two simplifications of the Navier-Stokes equation for
an incompressible fluid at low Re [Eq.\ (\ref{eq:lowRe})].
First, if the velocity profile $v$ takes the form of a so-called
potential flow $v=\nabla \phi$, with $\nabla^2\phi=0$, then 
Eq.\ (\ref{eq:incompressible1}) is
automatically satisfied. In this case,
solving the Laplace equation for a scalar field presents a vast simplification 
over the Navier-Stokes equation. For example, this approach allows us to
prove Stokes' Law, namely, that the speed $v$ at which a sphere of radius $r$ 
will tend to be towed in a Newtonian fluid with viscosity $\mu$ under 
the application of a force $F$ is given, in the limit of small Re, by
\begin{equation}
F = 6\pi\mu r v\;.
\label{eq:StokesLaw}
\end{equation}
Importantly, this approach can also be used to derive the theory of 
slender bodies, which underlies most of our understanding of undulatory 
physics at low Re. 

A more general simplification occurs in 2D. Then we may always write
$v = (\partial U/\partial y ,\, -\partial U/\partial x )$ 
where $U$ is a scalar potential. Hence 
Eq.\ (\ref{eq:incompressible1}) reduces
to the biharmonic equation
\begin{equation}
\nabla^4 U = 0\;.
\label{eq:biharmonic}
\end{equation}
This, in addition to the no-slip condition, gives a full description 
of low Re (incompressible) fluid dynamics in 2D.

\subsubsection{Slender body theory}
\label{sec:long_slender}
 
Undulations are particularly appealing to study, not only because of their
ubiquity but also because the motion can be elegantly formulated as the
propagation of a wave. In particular, in low Re it turns out that
some very general statements can be made within what is often dubbed
slender body theory.  We owe our understanding of the physics
of undulatory locomotion in large part to the pioneering works of physicists
and applied mathematicians (such as G.\ I.\ Taylor, M.\ J.\ Lighthill and 
G.\ J.\ Hancock) as well as zoologists (notably J.\ Gray, 
H.\ W.\ Lissmann and H.\ R.\ Wallace) in the 1950s
\cite{taylor51,taylor52,hancock53,gray53,wallace58,wallace59,wallace60}.  These
authors described the locomotion, analyzed the physical forces and derived the
mathematical framework that we still use today to understand low Re undulatory
locomotion. (Incidentally, the theory developed for high Re undulations bears
many similarities to slender body theory and is called elongated body theory 
\cite{lighthill71}).
In this section, we present a brief overview of key results for slender body
locomotion.

As we introduced above, two key asymmetries are required for an organism to be
capable of low Re swimming. Not only must the undulation be asymmetric under
time reversal, but some asymmetry in the environmental resistance is also
required (the latter being a more general requirement of locomotion at any Re).
Organisms that use undulatory locomotion are generally long and thin;
in fluids, this guarantees asymmetry in the resistance to forward (or backward) 
and sideways motion. 

Now as already noted, analytically solving the motion of non-spherical shapes
in a fluid is non-trivial. Slender body theory approaches this by deriving
approximate solutions of the Navier-Stokes equation for no-slip boundary
conditions applied to long cylindrical or similarly elongated shapes. These
solutions take the form of force-velocity relations.  Decomposing those into
their vector components then leads to two different linear force-velocity
relations along the major and minor axis of the object. It then becomes
possible to write drag equations [analogous to Eq.\ (\ref{eq:StokesLaw})] as
\begin{equation}
F_i = - c_i\,v_i\;,
\label{Stokes_drag}
\end{equation} 
where $c_i$ are the effective drag coefficients for motion tangent
($c_{\parallel}$) and normal ($c_{\perp}$) to 
the local body surface.

Within this framework, R.\ G.\ Cox was able to derive equations approximating
$c_{\parallel}$ and $c_{\perp}$ as functions of the length and radius of the
body, and the viscosity of the Newtonian fluid \cite{cox70}.  Approximating the body shape as a
prolate ellipsoid, J.\ Lighthill obtained similar but slightly more accurate
expressions for the effective drag coefficients \cite{lighthill76}: 
\begin{align} 
c_{\perp} &= L\,\frac{4 \pi \mu}{\rm{ln}(2q/r)+0.5}   \notag \\
c_{\parallel} &= L\, \frac{2 \pi \mu}{\rm{ln}(2q/r)} \;, 
\label{eq_cnct} 
\end{align}
where $L$ is the body length, $r$ is the body radius, $q=0.09\lambda$ (with
$\lambda$ being the wavelength of the undulation) and $\mu$ is the viscosity of
the fluid. The requirement of asymmetric drag forces can be neatly expressed in
terms of the ratio $K = c_{\perp}/c_{\parallel}$ which must have a value other
than unity if progress is to be possible. Notice that the viscosity of the
fluid has no effect on this ratio. Rather, it is completely determined
by the geometry of the object. For worm-like shapes, $K$ typically takes
values around $1.5$, whereas for infinitely long cylinders, $K$ approaches 2
($K \rightarrow 2$). 

\subsubsection{Undulations in rigid channels}
\label{sec:rigid}

To understand how propulsion is generated, we consider the simple case of a
cylindrical organism whose body undulates sinusoidally in a plane.  From the
scallop theorem we know that a standing wave will be unable to propel the
organism, so instead consider a traveling wave that is propagated backward from
head to tail, and is therefore not time symmetric. In the absence
of any fluid or walls (i.e., in vacuum), the body wave could still propagate
backwards at velocity $v_{\rm{wave}}$, but the organism would remain stationary.
By contrast, now consider the limit $K \rightarrow \infty$, which could be achieved 
by placing the organism in a tightly fitting sinusoidal channel with the same 
amplitude and wavelength as the body wave but with rigid walls \cite{gray53}.  
Within this channel, the wave is by definition stationary in global coordinates, 
forcing the organism forwards at a velocity $v_{\rm prog} = -v_{\rm wave}$.  

Let us examine how propulsion is achieved down the channel. As the wave is
propagated, the channel will apply a reactive force sufficient to prevent 
any motion in the normal direction. These 
reaction forces will only occur on the leading (i.e., backwards facing) edge 
of the wave. Now, the forwards directed components of these forces will add 
up, yielding a net propulsive force down the channel while any sideways 
directed components will cancel out.  Hence, the organism will move 
forwards through the channel, with some velocity $v_{\rm prog}$. 
As the organism slides forwards, it will rub against the sides of the channel 
and evoke reactive drag forces. Again the sideways directed 
components will cancel out over a cycle, but the backwards directed 
components will sum, yielding a net retarding force opposite to the 
direction of motion.  The propulsive force exerted by the walls of the 
channel will be exactly sufficient to counteract this retarding force when 
the organism progresses at velocity $v_{\rm prog} = -v_{\rm wave}$. 

\subsubsection{Low Re undulations: ``slip'' formulation}
\label{sec:slip}

\begin{figure}[htb]
\begin{center}
\includegraphics[width = 0.44\textwidth]{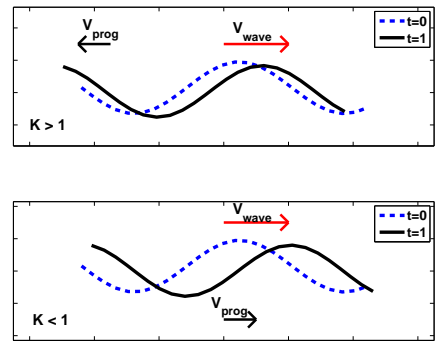}
\caption{The effect of $K$ on propulsion. An undulating body propagates a
sinusoidal wave to the right at velocity $v_{\rm wave}$ (red arrow). If
$K = 1$, no motion will result (not shown). If $K >1$, the body moves at
velocity $v_{\rm prog}$ in the direction opposite to $v_{\rm wave}$, 
with $|v_{\rm prog}| \le |v_{\rm wave}|$.  If $K < 1$, the
situation reverses and the body moves in the same direction as $v_{\rm
wave}$, with $v_{\rm prog}< v_{\rm wave}$.}
\label{fig:K}
\end{center}
\end{figure}

\begin{figure}[htb]
\begin{center}
\includegraphics[width = \columnwidth]{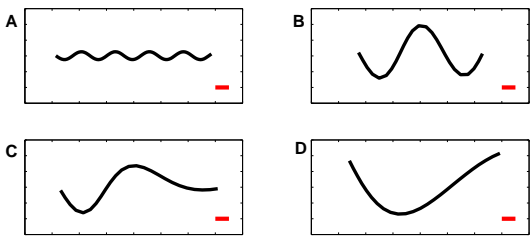}
\caption{Assessing the validity of Eq.\ (\ref{eq:slip}) with a physics
simulator (see text for details). A. The analytic result
of Gray {\it et al.} \cite{gray64} is an excellent approximation for an 
undulator with a
perfectly sinusoidal wave, a wavelength that is short relative to the 
body length and of a low amplitude. B. A purely sinusoidal idealization of the 
{\it C.\ elegans} crawling wave. The same equation handles such a wave 
reasonably well but with small errors due to the increased amplitude and 
wavelength. C. A more realistic {\it C.\ elegans} crawling waveform in which 
the wavelength increases towards the tail introduces significant errors.
D. A realistic {\it C.\ elegans} swimming waveform in which the wavelength 
is greater than the body length is very poorly described by the analytic
approximation.}
\label{fig:gray}
\end{center}
\end{figure}

Clearly, the $K \rightarrow \infty$ case can be trivially solved without
recourse to fluid dynamics or slender-body theory. Consider the same
organism in a fluid, i.e., with finite $K$ (the same reasoning will apply for
any $K > 1$).  Now, the normal component of velocity $v_{\perp}$ must
be non-zero if the normal force $F_{\perp}$ is to be non-zero. So, rather than
the wave remaining stationary in the global frame, it will slip backwards at
some velocity $v_{\rm{slip}}$ (with $|v_{\rm{slip}}| \le |v_{\rm{wave}}|$) while
the organism moves forwards at velocity $v_{\rm prog} = -(v_{\rm wave} - v_{\rm
slip})$ . This is schematically illustrated in Fig.\ \ref{fig:K}A. 
An approximation of slender body theory can be obtained from the so
called force resistivity theory and is due to Gray and Hancock 
\cite{gray55}. Here, rather than solving the Navier-Stokes equations 
for a long and slender body, the forces are approximated independently 
at each point (using equations of the form $F_i = c_i\,v_i$ as before)
and then integrated over the body length. Thus, any correlations in the 
fluid due to the spatially extended nature of the body are neglected.  
Applying this formalism to a perfectly sinusoidal body wave of low 
amplitude ($A$) and short wavelength ($\lambda \ll L$) (Fig.\ 
\ref{fig:gray}A), Gray {\it et al.} \cite{gray55,gray64} were able to 
derive an expression relating $v_{\rm{slip}}$ 
to $v_{\rm{wave}}$:
\begin{equation} 
v_{\rm{slip}} = v_{\rm{wave}} \frac{B + 1}{KB + 1} \; ,
\label{eq:slip}
\end{equation} 
where $B = 2 \pi^2 A^2 / \lambda^2$.  

In general, for a given locomotion waveform, the degree of slip depends only 
on $K$. Relating this back to the case of locomotion in a rigid channel where we
effectively have $K \rightarrow \infty$, we can see that for any $B$, we will
have $v_{\rm{slip}} \rightarrow 0$, so that over a single period of undulation 
the body travels a distance of one wavelength. Interestingly, slender body 
theory is valid for
any $K>0$.  For $K=1$ we obtain $100\%$ slip (equivalent to a vacuum, or
completely isotropic environmental resistance) and for $K<1$ we observe direct
wave propagation, with the body progressing in the same direction as the wave
(Fig.\ \ref{fig:K}B). 

While approximate, the simplicity and tractability of the force resistivity
theory has led to its extensive application in biological domains, in
particular for the study of flagellar propulsion in viscous fluids and
nematode locomotion in viscous and visco-elastic fluids (see below).  In both cases,
comparisons either against slender body theory \cite{johnson79} or against data
\cite{hfsp} have concluded that the approximations are reasonable.

\subsubsection{Beyond Newtonian fluids}

It is remarkable how adept biological organisms are at adapting to different
environments and modulating their behavior. Many organisms exhibit enormous
flexibility in navigating a wide range of environments, whether this involves
changes of gait or continuous modulation of a single behavior.  At the same
time, there are conditions in which organisms display very uncoordinated
locomotion or else fail to make progress altogether. These may correspond to
environments that are not usually encountered in an organism's natural habitat,
or -- more often in the geneticists' laboratory -- to mutants that lack an
essential protein. The investigation of biological forms of undulatory
locomotion across different physical environments dates back to the early 
20$^{\rm th}$
century \cite{looss11,gray64,wallace58,wallace59,wallace60} and is playing an
increasingly important role in genetics and in neuroscience 
\cite{horner08,lockery08,hfsp,shimomura08}. 

Until now, we have limited ourselves to Newtonian fluids, which are fully
described by the viscosity of the medium, but many low Re swimmers in
fact move through complex fluids or other `soft' environments. 
For slender bodies in a Newtonian fluid, the ratio of drag coefficients 
$K$ is fully determined by the geometry and cannot exceed 2. In contrast, 
in non-Newtonian environments, $K$ (if and when it is well defined) is 
both a function of the geometry and of the medium.
Strictly speaking, viscosity is not defined in non-Newtonian fluids, since
the linearity of Eq.\ (\ref{eq:viscosity}) is violated. Thus even the 
Navier-Stokes equation is not applicable. 
For example, the
fluid may have some non-trivial structure, it may have energy storage
capacity (e.g., elasticity), or perhaps the properties of the fluid may
depend on the speed with which it is deformed.  Of these cases, visco-elastic 
fluids (and visco-elastic approximations of gels) are probably the most 
relevant to biological swimmers. Slender body or resistive force
theories can both be straightforwardly extrapolated to model visco-elasticity
when the elastic properties of the fluid can be approximated by effectively 
stronger resistive drag coefficients in the normal direction, thus 
increasing the ratio $K$ 
\cite{gray64}$^,$\footnote{Note that visco-elasticity is typically modeled by
adding time-dependent terms to the Navier-Stokes equation, so this
parametrization by $K$ is only valid in the limit where this time dependence 
can be neglected.}. 
In what follows, we loosely call such extensions to Newtonian
fluids, {\it anomalous $K$ environments}.

To the extent that slender body theory (and the simplified slip formulation
thereof) may be extrapolated in this way, all of the above logic still holds, 
except that now the slippage parameter $K$ is no longer determined solely by 
the geometry of the swimmer.

\section{Applications to biological swimmers and crawlers}

The hypothetical cylindrical organism described above is a good approximation
for a vast range of different micro-swimmers, from bacteria to
microscopic larvae or worms. Of these, the best studied examples are
flagellated microorganisms. Indeed, this field already occupies a vast
literature (see \cite{morgan01,ginger08,sowa08} for reviews). Here we present only a brief overview of
two examples of molecularly driven undulations at the subcellular level: 
flagella mediated swimming in spermatozoa and in bacterial cells. We then
proceed to discuss locomotion in non-Newtonian environments, using small 
nematode worms as a case study.

\subsection{Flagellated microswimmers}

Bacteria and spermatozoa self-propel by means of organelles called
flagella. A flagellum is a long filament that protrudes from the cell and
provides propulsion. However, the flagellar motion of eukaryotic cells
(like sperm) is significantly different from that of bacteria. While a
sperm's flagellum makes sinusoidal undulations in 2D, bacterial flagella
make helical undulations. 

Eukaryotic flagella consist of a core of two microtubules (long, tube shaped
polymers) surrounded by nine microtubule doublets, arranged to form a circle.
Each microtubule is as long as the entire flagellum and typically extends a 
few body lengths. The flagellum of, for example, a sea urchin spermatozoon 
undulates at about 50Hz, with a nearly sinusoidal waveform \cite{eshel90}. 
This bending is achieved by what is
essentially a distributed molecular motor consisting of the microtubules
themselves. Powered by the hydrolysis of ATP, pairs of microtubule doublets
slide relative to each other \cite{loewy91}. In order to generate the alternate 
bending characteristic of a sine wave, this sliding motion is resisted by local 
inter-doublet links at specific points along the flagellum. 
By propagating these zones of
sliding and linking along the flagellum, the sinusoidal wave is generated and
propagated. The control of this process is chemically ``hard wired'' into the
system.

Despite generating force according to similar underlying physics, the control of
a bacterial flagellum is completely different \cite{loewy91,ginger08} and unique 
to bacteria.  Unlike eukaryotic flagella, bacterial flagella are passive fibers 
incapable of active bending. The solution they use instead involves helical waves.  
The bacterial flagellum itself is a thin filament, grossly similar to the 
eukaryotic one. However, rather than the microtubules of eukaryotic flagella, 
bacterial flagella consist of proteins called flagellin arranged to form a hollow
cylinder. Subtle properties of the flagellin's physical structure mean that the
resulting flagellum is not straight but helical, which is integral to their
function in locomotion. The entire helical flagellum is connected via a
sharply bent construct called a hook to the shaft of a rotational motor in the 
cell's membrane \cite{sowa08}. This molecular motor bears a striking resemblance 
to an electrical stepper motor and is powered by the flow of protons across the
cell's membrane. Thus the flagellum acts exactly like a boat's propeller
attached to the drive shaft of a motor. 

How does this mechanism generate thrust?  In the case of eukaryotic flagella,
the posteriorly directed components of the motion add up but the sideways components
counter balance, thus generating progress in a direction parallel to the long
axis of body (axially) as described in Section \ref{sec:slip}.  Now, to understand 
bacterial flagella, think of each part of the helix as a small tilted rod 
\cite{taylor67}.  When the rod rotates within the spiraling helix, it exerts an 
axial and sideways force on the fluid leading to forward propulsion and an overall 
torque. The torque then induces a counter rotation of the cell body. 

On a final note, we mention the interaction that occurs when multiple flagella
surround a bacterial body and interact to generate complex motion.
This relies on the fact that the motor driving each filament can spin 
both clockwise and counterclockwise.  When spinning counterclockwise, all the 
flagella rotate synchronously, forming a bundle that propels the cell forwards. 
When spinning clockwise, however, the filaments fail to synchronize and each
one ends up pushing the cell in a different direction
\cite{berg96,macnab77}. Thus the bacterium goes nowhere,
but does undergo some random rotation. 

On a final note, we mention the interaction that occurs when multiple flagella
surround a bacterial body and interact to generate complex motion.
This relies on the fact that the motor driving each filament can spin
both clockwise and counterclockwise.  When spinning counterclockwise, all the
flagella rotate synchronously, forming a bundle that propels the cell forwards.
When spinning clockwise, however, the filaments fail to synchronize and each
one ends up pushing the cell in a different direction
\cite{berg96,macnab77}. Thus the bacterium goes nowhere,
but does undergo some random rotation.
In a typical environment, a bacterium will switch between the two modes of
flagellar rotation and alternately rotate and swim in a so-called {\em
tumble and run} motion. Bacterial sensory mechanisms modulate the frequency
of tumbles, such that when a bacterium is in a favorable environment (rich
in nutrients) it will turn often and tend to stay in the same vicinity, whereas
if conditions are less favorable, runs will be longer and the bacterium will
undergo a biased random walk in search of food \cite{vicsek}.

\subsection{Undulations of microscopic worms}
\label{sec:groove}
To demonstrate some of the physics of simple non-Newtonian
environments, we focus in what follows on a small nematode worm called 
{\it Caenorhabditis elegans}. {\it C.\ elegans} is a leading model
organism for biologists and as such is grown extensively in laboratories,
where it is cultured on the surface of agar gels. With a length of 1mm
and up to 2Hz undulation frequency, the worm can have a Re of about 1 in
water\,--\,approaching the upper limit but still within bounds
for a low Re treatment. 

{\it C.\ elegans} worms crawl on the agar surface with very little slip, 
suggesting that the value of $K$
is high. Close inspection reveals sinusoidal tracks left by the worms as
they move. 
%(Fig.\ (\ref{fig:groove}). 
Each track is actually an indentation 
(or groove) in the surface of the gel, and helps to explain the lack of slip. 
Although the worm's mass is negligible, a thin film of water forms around it
and the resulting surface tension presses it strongly against the gel surface 
\cite{wallace68}.
As the worm moves forwards, it overcomes the gel's yield stress and breaks the
polymer network with its head. This allows the rest of the body to slip
forwards more easily. Motion normal to the body
surface is strongly resisted because such motion would require further breaking
of the polymer network, over a much larger area. For a sufficiently stiff gel, 
the net result is functionally quite similar to locomotion in a solid 
channel and allows $K$ to be significantly larger than in the Newtonian case
\cite{gray64,niebur91}. 

A similar effect occurs when an organism moves through a granular medium like
soil. Again the head acts as a wedge, forming a channel through the grains
\cite{wallace58,gray64}. Motion in the normal direction would require many more
grains to be displaced.  Although the resistive forces of these media cannot
strictly be called drag, their behavior is nonetheless frequently represented
in terms of local resistance coefficients $c_{\parallel}$ and $c_{\perp}$. In
many cases the non-Newtonian properties of these fluids are sufficiently well
modeled using this approach. 

\subsubsection*{Experiments}

Given the fundamental importance of environmental properties to undulatory
locomotion, it would be beneficial to experimentally determine the relevant 
properties of the swimmer's environment (and in the case of the worm, 
particularly those of agar gels). This would help to determine how valid the 
slip formalism is for such media and to facilitate the development of 
quantitative models of an organism's locomotion. 
For an ideal Newtonian fluid it suffices to know the viscosity, which can be 
easily measured by even the simplest rheometers.  Given the fluid viscosity and
the dimensions of the organism, good estimates of the drag coefficients
can be obtained. If greater accuracy is required then the full Navier-Stokes 
equations may be used.  However, even in Newtonian environments,
complications can arise. For example, nonuniformity in the medium may result 
in inhomogeneous viscosity. This is especially a problem for very small
swimmers, where inhomogeneities on scales of microns or tens of microns may
locally influence the motion.  Additional complications arise when
dealing with visco-elastic fluids, gels and suspensions of particles. While
even complex fluids can be well characterized using a modern rheometer,
the quantities measured (e.g., the elastic and loss moduli as a function of
frequency) are not behaviorally relevant. In principle, advanced
simulation techniques could be applied to predict physical forces from
rheological properties, but this would be an extremely difficult
undertaking. 

It turns out that a simple representation of fluid behavior in terms of local
drag coefficients $c_{\parallel}$ and $c_{\perp}$ is sufficiently accurate for
many types of study, if only we could measure their values. One of the
complicating factors in such an experiment is the small size of the organisms
under investigation. Nonetheless, some innovative approaches have been used. 

By dropping small diameter wires through various Newtonian fluids, Gray and
Lissmann \cite{gray64} were able to verify the expected value of $K
\approx 1.5$. By performing the same experiment in agar gel, they were able to
show that $K$ was significantly greater (though no value is reported). In order
to quantitatively estimate the forces experienced by a small nematode crawling
on agar, Wallace placed small glass fibers (with dimensions similar to
the nematode) on the agar surface and measured the force required to pull them
along, thus estimating the propulsive force exerted by the nematode
\cite{wallace69}. He also performed experiments where the thickness of the 
water film around the worm was altered, thereby modulating the forces acting 
on it and resulting in different locomotory behavior. More recently, Lockery 
{\it et al.} \cite{lockery08} have developed an alternative strategy of imposing 
$K = \infty$ by manufacturing micro-fluidic chips with tiny channels of
pre-determined shape filled with water.  
While direct measurement of physical
forces in fluid environments would be preferable, the ability to effectively
set these parameters to known values is a powerful tool.

\section{Simulations}
\label{sec:worm_sims}

Undulatory locomotion of roughly cylindrical organisms at low Reynolds
number is well suited to implementation in computer simulations. In particular,
when body bending only occurs in a plane, as in {\it C.\ elegans}, 
a 2D simulation is sufficient to capture the dynamics, making the task 
simpler and less computationally expensive. In what follows we describe
a number of simulators that can be used to study worm locomotion in a variety
of media.

\subsection{$K$ estimation}

As we have already seen, in low Re environments, the trajectory of a body
is entirely determined by its sequence of shapes and the properties of
the environment. One approach then, is to use the locomotion traces
of different bodies to estimate the properties of the environment.
If the model of the environment is approximate, such as reducing the 
description of the environment to two drag coefficients, then a simulation 
approach can also serve to assess the validity of the theory \cite{hfsp}. If 
valid, the simulator can then be used to assess further approximations, such
as the slip formalism in Eq.\ (\ref{eq:slip}) \cite{gray55,gray64}.

\subsubsection*{Solving equations of motion at low Re}

In a typical, high Reynolds number physics engine, the net force on a body
results in acceleration according to $a=F/m$. At low Re, the
very small mass will lead to very large accelerations, leading to a stiff
system requiring very short time steps. In the limit of $m \rightarrow 0$
and for $F \neq 0$, we will have the numerically problematic situation of
$a \rightarrow \infty$. In the ``real world'', this means that the velocity
of the body will always be at steady state, at which the net force and
similarly net torque are zero. Solving the equations of motion is therefore
tantamount to satisfying these conditions.

For example, one might begin by comparing the body shapes at times $t(i)$ and
$t(i-1)$, assuming zero progress. Given the model of the environment, one
can then calculate the reactive environmental force and combine these to obtain
the net force ${F}_{\rm{net}}$ and torque ${{\tau}}_{\rm{net}}$ acting on 
the body.  For our simplified model of Newtonian or anomalous $K$ 
environments, one need simply decompose the velocities of each point into their
normal and parallel components and calculate the corresponding drag forces.
Based on the directions and magnitudes of these vectors, we can now apply a
small displacement and/or rotation (in the direction of $-{F}_{\rm{net}}$ and
$-{{\tau}}_{\rm{net}}$) and recompute the reactive forces to get the new resultant.
This process can be iterated, accumulating small displacements/rotations
until the zero net force and torque conditions are met (within a specified
tolerance), at which point we combine the final 
displacement and rotation
$\delta x$, $\delta y$ and $\delta \theta$ 
with the center of mass trajectory. Thus we are effectively 
performing a gradient descent on ${F}_{\rm{net}}$ and ${{\tau}}_{\rm{net}}$ 
in order to find the correct trajectory. 

Within the slip formalism, the progress of the organism does
not depend on the actual values of $c_{\parallel}$ and $c_{\perp}$, just on the
ratio $K$. Given that internal forces are not modeled, it is simply
assumed that they are sufficient to produce the recorded changes in body
shape. Thus, for a known $K$, this kind of simulator can simply solve
the equations of motion. Alternatively, if $K$ is not known, but the 
coordinates and rotation of the body shapes are known, then 
a simulator like this can also be used to estimate $K$. Thus, this
simulation approach can be used to estimate environmental properties
without any direct measurement of that environment (except the visual
recording of the motion) \cite{hfsp}.

\subsection{The physics simulator applied}

As mentioned above, a physics simulator is particularly valuable for 
computing the progress of a long and slender undulator in a Newtonian
or anomalous $K$ environment. In addition, such a tool can also be used 
to address basic questions in slender body theory or in low Re physics.
Here we briefly describe three such examples.

\subsubsection{Limits of the slip formulation}

The physics simulator as described above uses a basic result of slender
body theory, namely that motion lateral and tangential to the body experience
different drag forces (and extrapolated to arbitrary ratios of these forces).
However, it makes no assumptions about the configuration of the body. One can
therefore ask \cite{hfsp} (i) whether the results of slender-body physics are 
valid approximations of, say, the forces experienced by worms moving on agar 
and (ii) whether further simplifications, e.g., as given by Eq.\ (\ref{eq:slip}), 
are valid. The latter
would require a simulation of artificially generated body shapes in a
pre-specified environment and the comparison of the simulated progress with the
theoretical prediction.  For example, Fig.\ \ref{fig:gray} shows that three 
of the
assumptions needed to derive Eq.\ (\ref{eq:slip}) break down for worm-realistic
skeletons. These are first, that the locomotion wave is sinusoidal; second, that
the wavelength $\lambda$ is short compared to the body length $L$; and third,
that the amplitude $A$ is sufficiently small so that the wavelength $\lambda$ 
is similar to the corresponding arclength along the undulating object's body.
Such results, while straight forward to generate, nonetheless
offer a quantitative handle on commonly made approximations in the field.

\subsubsection{Purcell's scallop in non-Newtonian environments}

One interesting question is to what extent the scallop theorem applies
to more complex environments. Reviewing the argument in Sec.\ 
\ref{sec:scallop}, it is easy to see that the extension of the physics
to variable $K$ does not introduce any time asymmetry into the governing
equations, and so the scallop theorem should hold. The physics simulator
above is well suited to simulating scallops in media with different values
of $K$, so this generalization of the scallop theorem can easily be 
verified numerically and indeed holds true, as would be expected. 

\subsubsection{The variational principle and Helmholtz's Theorem}

As described above, the conventional way to determine the dynamics of objects
in Newtonian fluids requires a solution of the Navier-Stokes equations, but
this can prove to be computationally difficult. To complement this approach, attempts
have been made to tackle fluid dynamics, or at least low Reynolds number
Newtonian fluid dynamics, from a very different perspective: that of
minimization of energy dissipation. 
Put simply, if the evolution of the system is too difficult to calculate, one
can put forward an arbitrary path, and then use variational methods to find the
path that minimizes the energy dissipation of the system, be it heat
dissipation in an electrical circuit \cite{Kirchhoff} or at least in
principle, due to drag forces in a fluid \cite{helmholtz59}. If this so-called
{\it principle of minimum energy dissipation} was true, this path would then 
satisfy the
equations of motion. In fact, minimum energy dissipation has been proved
anecdotally, in very simple cases, but in other cases it has been found not to
hold at all. Thus, while appealing and intuitive, it appears not to be a
general principle.  Having said this, it is still an open question to try to
understand under what conditions the principle applies, typically because of
the potential for computational advances in difficult problems where the
equations of motion are particularly challenging to solve.

What then, is the status of this problem in fluid dynamics?  A theorem due
to Helmholtz and Korteweg proved the minimum energy dissipation
principle in the case of a bounded volume filled with a viscous fluid
where the boundaries are moving with a well specified velocity, and in the
limit of negligible inertial forces \cite{helmholtz68,korteweg83}. Only very
few studies have attempted to tackle other boundary conditions, such as
spheres or ellipsoids immersed in a Newtonian fluid and moving under
gravity \cite{jeffrey22,taylor23,christopherson55,dowson59}. The
more general case remains open. 

One interesting case for which results have not been derived is
that of a self-propelled body, i.e., with a time-changing configuration.
Another interesting extension is to non-Newtonian media.
We restrict the discussion to undulatory swimming through Newtonian and
anomalous $K$ fluids at low Reynolds number, where despite being intractable 
analytically, it is possible to study the problem using numerical models 
and simulations.
An advantage of a simulation approach is that it can be applied to arbitrary 
artificial body skeletons, as well as body shapes extracted from movies of 
actual swimming (where both the shape and coordinates of the body are known) 
and for a variety of (Newtonian as well as non-Newtonian) drag coefficients.
For a long and slender body in low Re, motion in any
environment that can be described by drag coefficients $c_{\parallel}$ and
$c_{\perp}$ will be subject to energy dissipation of the form
\begin{equation} 
E = \int_0^T \int [ v_{\perp}^2(t,s) c_{\perp} + 
                      v_{\parallel}^2(t,s) c_{\parallel} ]\, ds \, dt 
\label{eq_energy} 
\end{equation} 
over a cycle time $T$,
where $v_{\perp}(t,s)$ is the normal component of velocity at time $t$ and at
position $s$ on the body surface, and $v_{\parallel}(t,s)$ is the corresponding
tangential component.  Equation (\ref{eq_energy}) guarantees only non-negative 
contributions to energy dissipation, and easily lends itself to numerical 
minimization. 

\begin{figure*}[htb]
\begin{center}
\includegraphics[width = 0.95\textwidth]{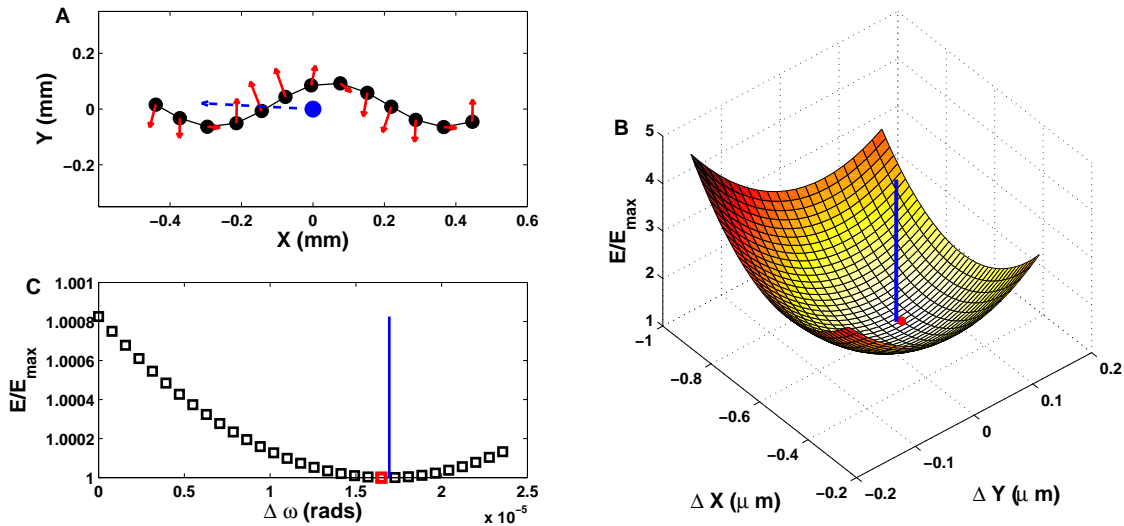}
\caption{The principal of minimum energy dissipation correctly predicts the
displacement and rotation of a self-propelled worm in a 2D anomalous $K$
medium. Simulations confirm that the solution of the equations of motion
coincides with the minimum energy dissipation.  A. An undulating body with
sinusoidal shape moving according to the equations of motion.  Red arrows
represent the drag force acting on each point. The ``CoM" (blue circle) moves
with velocity V (blue arrow) such that the net drag force is zero.  B, C. The
corresponding energy dissipation landscape based on a sweep of displacements
(x,y) in a plane (B) and rotation in the plane (C). In both plots the minimum
energy dissipation is marked in red on the landscape. The motion obtained from
the equations of motion (shown in A) is indicated by a blue line. The results
agree to within the step size.}
\label{fig:dissipation}
\end{center}
\end{figure*}

Figure \ref{fig:dissipation} demonstrates this approach with one such 
example. The body and environment in this example are represented in the
same way as described in Sec.\ \ref{sec:long_slender}. For this example (and 
indeed for
a variety of waveforms and actual movies of {\it C.\ elegans} locomotion),
the minimum energy dissipation trajectory agrees with the solution of
the equations of motion, to within the discretization step of the energy 
sweep. The simulation also demonstrates the intuitive result that 
the landscape (as far as explored) is smooth and has only a single minimum.
This approach, while far from an analytic proof, 
demonstrates the potential application of such simulators. Specifically, 
the set of simulations in this example suggests that slender body motion, 
whether passive or self-propelled, in any environment characterized by 
$c_{\parallel}$ and $c_{\perp}$, will obey the principle of minimum energy 
dissipation. 

\subsection{Modeling worms}
\label{sec:worms1}

A physics simulator like the one described above is useful in cases where
the locomotion waveform can be recorded or pre-determined, but intentionally
avoids the question of how an object or organism generates its undulation 
wave.  In general, the shape of the body is determined by a combination
of (i) internal (molecular or muscle) actuation forces, (ii) physical 
properties of the body and (iii) external forces from the environment. 

Consider again our friend {\it C.\ elegans}.  Conveniently, the worm's neuro-anatomy
essentially limits it to bending in two dimensions \cite{white86,chen06}.  Nematodes, like other
worms, lack any form of rigid skeleton and instead rely on the antagonistic
forces of their elastic outer casing (cuticle) and the hydrostatic pressure
within their body cavity to maintain their shape.  Thus, at the simplest
level, the body can be represented by an elastic rod or cylinder with
some internally controlled bending along the rod. Alternatively, more
elaborate and detailed models are possible that allow for more accurate
biophysical and biological grounding. 

In the nematode literature, the canonical example is a model of {\it 
C.\ elegans} locomotion due to Niebur and Erd\"os \cite{niebur91}, which is based
on exactly these principles.  Their 2D model approximates the body as an 
elongated rectangle represented by two rows of points along its outline. 
Adjacent points are connected by springs (representing the
cuticle) and are pushed apart by pressure forces. 
This body also behaves effectively as an elastic rod with a characteristic
persistence length. Note that until now, we have not evoked any low Re
considerations (except that we have not endowed our `rod' with any mass).

The low Re physics does come in when solving for the changes of shape
and position of the body.  In the above model, points are able to move
independently under the action of muscle forces, and modulated by the
environment. To solve the model, one can invoke once more the slender body
theory and approximations due to Gray {\it et al.} \cite{gray55,gray64}, 
and decompose the forces 
into tangential and normal components $F_{\parallel}$ and $F_{\perp}$.
Much like the physics simulator described above, the zero net force 
condition can be applied to calculate the velocity of each point along the body.
Specifically, we can write write 
$F_{\rm{drag},i} = -c_i v_i = - F_{\rm{internal},i}$ 
along each direction $i$, and since we already have our internal forces, it 
remains to extract the speeds along each direction 
$v_i = F_{\rm{internal},i} / c_i$. 

\section{Sensory integration and closed loop control}

To complete the model of an undulating system, one may also include the active
elements that induce and control the bending of the body.  This may be the
molecular biophysics underlying force generation in flagellar motion, or a much
higher level model of brain and body control in an undulating animal.  Either
way, one can think of a motor control system as consisting of a control signal
that drives actuators or motor elements to control the shape of 
the body, subject to environmental forces. In animals, the control signal is
generated by the nervous system; the actuators are muscles; and feedback
messages are generated either by the body (signals that sense the state of the
muscles as well as the orientation and shape of the body) or by other sensory
perception (vision, etc.).  For example, when we lift a heavy object or push
against a wall, the body senses a strong resistance from the environment and
can relay this to the nervous system which, in turn, can alter or modulate the
neural control. Thus, in this example the neural signal can be thought of
essentially as a centrally controlled system with some feedback. It turns out
that the vast majority of animal motor control systems contain such neural
``control boxes''. Since most muscle behavior consists of rhythmic motion, 
the underlying neural control generates rhythmic patterns of activity. Neural
circuits that generate rhythmic patterns of activity even when completely
severed from the rest of the body are dubbed {\it central pattern generating}
or CPG circuits \cite{marder96,marder00,harriswarrick93}.

The fact that a CPG functions in isolation suggests that it can also be 
modeled in isolation. Indeed, traditionally, most animal motor control 
models (locomotion included) tended to be limited to a bottom-up 
model of internal neuronal or neuro-muscular dynamics. These models 
are then complemented by top-down models of the physical aspects of
locomotion (e.g., the aerodynamics of flight, the mechanics of legged 
locomotion and so on). The underlying assumption here is that, to a first 
approximation, the neuronal control can be treated as a stand-alone control 
unit and hence decoupled from the physics of the body and environment. 

When then does the physics matter sufficiently to justify an integrated
neuro-mechanical model?  In what follows, we describe two examples of 
swimming undulators: the lamprey, and {\it C.\ elegans}. In the lamprey,
the vast majority of modeling has focused entirely on the isolated neural 
control of locomotion. Recent exceptions are beginning to highlight
the role that physics plays in the locomotion and are perhaps opening
new avenues of investigation of this classic model system. In {\it C.\ elegans},
the low Re physics has long been recognized to play a significant role,
but models have still tended to decouple the nervous control from the 
mechanics. Here too, experiments and models are moving increasingly in
a direction of an integrated study of the entire locomotion system, 
bottom-up and top-down.

\subsection{A high Re example: Lamprey swimming}

The lamprey is a primitive, eel-like aquatic vertebrate that can 
reach up to one meter in length.  
It swims by propagating lateral undulations 
with increasing amplitude along its body (from head to tail).  Lampreys have 
a brain and spinal cord with a relatively complex nervous system.  The spinal 
cord serves to relay information between the brain and body, but is also capable 
of complex dynamic behavior.  In the 1980s it was shown that the basic neural
activity pattern responsible for locomotion could be produced by the isolated
spinal cord \cite{cohen80}, without any input from the brain or from sensory 
pathways.  This became the canonical example of a vertebrate CPG circuit 
\cite{cohen82,cohen87,kopell88,ermentrout94,grillner98}.

\begin{figure}[htb]
\begin{center}
\includegraphics[width = 0.8\columnwidth]{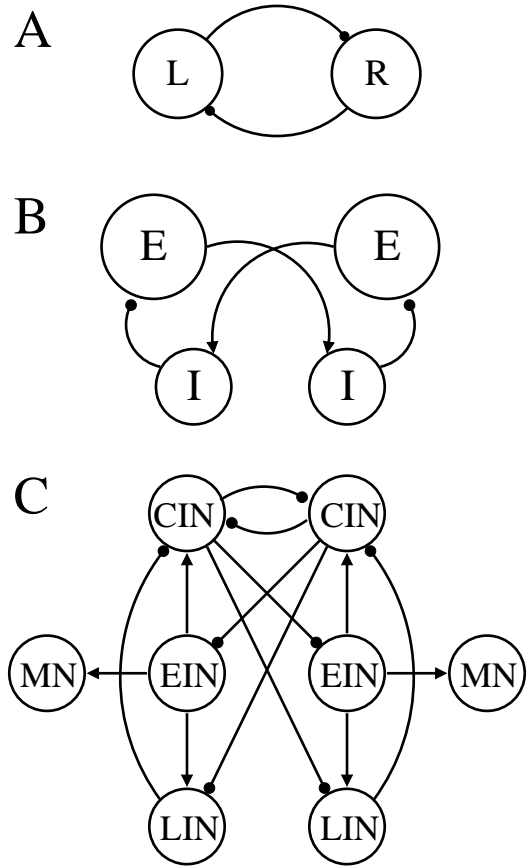}
\caption{Schematics of neural oscillators for motor control. 
A. The half-center oscillator is the simplest central pattern generating 
circuit (excluding intrinsically oscillating neurons) and consists of two 
mutually inhibiting neurons. An additional requirement is that the neurons 
can be released from, or are able to escape from, the inhibition. An example
of a release mechanism is synaptic fatigue, where the inhibition wears out 
over time. B. Simplified nematode oscillator network (adapted from 
Ref.\ [\protect{\onlinecite{stretton85}}]). 
% RESUBMIT: reference added
(E)xcitatory neurons 
innervate the (I)nhibitory neurons which in turn relay the inhibition to 
the opposing excitatory neuron. C. The lamprey segmental oscillator 
(adapted from Ref.\ [\protect{\onlinecite{ijspeert99}}]) is more complex, 
but functions along 
similar principles. It consists of motor neurons (MN) along with three 
classes of interneuron (IN), with CIN providing the contralateral 
inhibition.}
\label{fig:cpgs}
\end{center}
\end{figure}

It was further shown that the lamprey CPG consists of a sequence of
semi-independent neural oscillators that are capable of independent pattern 
generation, but are coupled to each other in such a way that the correct phase 
lag between adjacent units is preserved \cite{cohen87b}. Thus, the CPG circuit 
can generate stable rhythmic activity (at swimming frequencies) that propagates 
along the spinal cord, and where each `segment' along the body exhibits 
anti-phase oscillations between the left and right sides of the body (so 
that when connected to the muscles, one side would contract when the 
opposite side relaxes). In most CPG circuits, the prevalent model of anti-phase 
oscillators relies on the {\em half center oscillator} 
\cite{grillner98,churchland92}, which consists of two units that are
coupled via reciprocal inhibition (as shown schematically in Fig.\ 
\ref{fig:cpgs}A). These units typically represent separate pools of neurons 
(as they do in the lamprey).
% but minimally even two neurons suffice.  
External driving forces and/or internal 
dynamics modulate the activity to generate rhythmic, out of phase activity.

Early models of the lamprey spinal cord relied on exactly these principles, 
and have led to important advances, both in the study of the lamprey and other 
nervous systems \cite{cohen82,cohen92,ijspeert05b}, and in our understanding of 
coupled oscillator systems in nonlinear dynamics 
\cite{kopell86,kopell88,ermentrout94}.  While the lamprey swimming CPG circuit (Fig.\ 
\ref{fig:cpgs}C) looks rather complicated, the motif of reciprocal inhibition
that is required for generating oscillations is not very different 
from the minimal circuit in Fig.\ \ref{fig:cpgs}A. The circuit includes both excitatory
% RESUBMIT: sentence above modified
and inhibitory pools of neurons that innervate motor neurons that connect to the 
muscles.  Excitatory interneurons (EIN) provide the excitation to switch the circuit 
on; contralateral inhibitory interneurons (CIN) provide the cross inhibition,  
and lateral inhibitory interneurons (LIN) help to terminate contraction of each 
side \cite{ekeberg93}.  This principle is demonstrated in an elegant robotic 
instantiation by Auke Ijspeert {\it et al.} \cite{ijspeert05}, that builds on 
the integrated neuro-mechanical model of lamprey swimming due to Ekeberg 
\cite{ekeberg93}.

Over the years, as models of the lamprey motor control have become more refined 
and sophisticated, research has increasingly focused on modeling the modulations 
of the swimming rhythms by input from the brain, from sensory neurons and 
by the local action of chemicals (so-called neuromodulators) and even
modulation by proprioceptive inputs, but of these, the role of sensory effects 
are least understood \cite{cohen99}.
To shed some light on the effects of sensory feedback, 
Simoni and DeWeerth \cite{simoni07} have recently
introduced a neuromechanical lamprey model with sensory feedback to study 
synchronization effects between the CPG circuit and the mechanical resonance 
of the body. In principle, this approach, if further complemented by
an understanding of the fluid dynamics, should lead to further progress
in understanding how the lamprey and other systems exploit the physics
of the environment to modulate their internal neural dynamics and motor control.

\subsection{A low Re example: {\it C.\ elegans}}

Imagine a system in which the neural control is driven directly by sensory 
signals (such as reflexes). 
Such a neural system would completely fail in the absence of a body.
Surprisingly, such appears to be the minimal
model describing {\it C.\ elegans} locomotion \cite{niebur91,hfsp,boyle09}.  
We have already briefly described some of the worm's means of locomotion
and its exploitation of the environment to generate thrust (see Secs.\ 
\ref{sec:groove} and \ref{sec:worm_sims}), but here we focus on the internal 
(neural) control of the worms' undulations.  

This tiny nematode worm has an invariant anatomy that consists of a mere 959
cells in the adult.  Of these, about a third (302 cells) are nerve cells or
neurons. These make up a simple nervous system that is distributed throughout
the animal. Remarkably, the nervous system appears invariant. In other words,
the connectivity diagram or neural circuitry is the same across worms. This
connectivity diagram was meticulously mapped out \cite{white86,chen06} and is
the starting point for understanding the worm's behavior. 

Locomotion is the worm's principal motor activity and mediates everything the 
worm does, from foraging to mating. Thus,
it is not surprising that of its 302 neurons, around 200 are involved in
locomotion.  Of those, about 50 are directly responsible for the generation of
undulations and their backward propagation down the body, i.e., the generation
of forward locomotion. (The worm can also move backwards, propagating waves
from tail to head, but this behavior relies on a different set of neurons). To
put these numbers in perspective, the human brain contains of the order of
$10^{11}$ neurons and an insect such as the tiny fruitfly has about $10^5$.
Thus, while the nervous systems of most animals (including lamprey and even the
fruitfly) rely on large populations of neurons to perform processing, in
{\it C.\ elegans} single neurons must do the same job. 

When nematodes are placed in environments with different viscous, 
visco-elastic or gel properties, the parameters of their locomotion wave 
change quite significantly.  For instance on agar gels,
the worms exhibit slow 
($\sim$0.5Hz) short-wavelength undulations (with the body length spanning 
about 1.5 wavelengths), whereas in watery environments, they appear to thrash, 
i.e., undulating much more quickly (at $\sim$2Hz) with a wavelength of just 
under twice the body length. In fact, the entire range of intermediate
behaviors can also be obtained in environments with various degrees of
visco-elasticity \cite{hfsp}. In principle, the observed change in behavior 
could be the result of internal changes to the neural circuit (e.g., 
by neuromodulators, or sensory pathways that activate different sub-circuits). 
Alternatively, the entire change in behavior may be down to changes in the 
physical interaction between body and environment. 

To shed light on this problem requires a closer look at the locomotion
nervous circuit of the worm. Unlike vertebrates, the worm does not have
a spinal cord, but it does have an analogous ventral cord which contains
the locomotion motor neurons, and so we shall focus on it. First, one may 
look for evidence of characteristic motifs along the ventral cord that 
are expected to generate rhythmic activity, such as reciprocal inhibition. 
This typically requires breaking up the circuit into small units (analogous
to segments or vertebrae in the lamprey).  Unfortunately, even this first
step is not a trivial task, since the nervous circuit of the worm contains 
very different numbers of neurons on different sides of the body, and
so is not naturally divisible into repeating units.  Based in part on 
data from related nematodes, in which direct recording from neurons is 
possible, a model of the generic nematode segmental oscillator (Fig.\ 
\ref{fig:cpgs}B) was proposed \cite{stretton85}. This circuit bears a 
distinct resemblance to the simple
half-center oscillator in Fig.\ \ref{fig:cpgs}A, except that the inhibition 
is mediated via distinct inhibitory neurons. 
Nonetheless, for a combination of reasons, questions have been raised about
the ability of this circuit to support a CPG network. 

One may therefore ask what would be a working model of the locomotion
in the possible absence of a CPG mechanism. One interesting conjecture
is that the worm may use a sensory feedback mechanism as the primary 
driver of its neuronal oscillations. Weight is given to this hypothesis 
by several models of the worm's locomotion that are grounded in its 
anatomy and successfully produce oscillations by means of mechanosensory 
feedback \cite{bryden08,karbowski08}. How would such a mechanism work?
We begin by discussing the worm's so-called crawling behavior (short 
wavelength undulations observed on agar). 
This was addressed by the first model of {\it C.\ elegans} locomotion due
to Niebur and Erd\"os \cite{niebur91}. Their model of the worm's body was 
introduced in the Sec.\ \ref{sec:worms1}, but they also proposed a simple model 
nervous system. 

This model involved a hypothetical CPG circuit in the worm's head, allowing it
to generate a sinusoidal trajectory that was then preserved by strong lateral
resistance, due to a very stiff groove (high $K$). The worm's shape was
therefore determined by the shape of the groove, and the model nervous system
was then responsible for generating muscle contraction down the body, pushing
the worm forwards. To control the rhythmic muscle activity and rhythmic
underlying neuronal activity, the neurons in the model were endowed with
so-called {\it stretch receptors}, or mechano-sensitive ion channels that
activate an ionic current in response to stretch. The current then activates
that neuron, and local muscle contraction is induced.  If muscles on one side
of the body are maximally contracted, the tissue on the other side is maximally
stretched.  Crucially, the model assumes that the stretch receptor signal is
transmitted some distance forward from the site of the stretch: The existence
of a set distance between the stretch detection and the ensuing contraction can
result in the propagation of the undulation in the desired direction (from head
to tail).  This integrated neuromechanical model was able to reproduce the
crawling behavior on agar if the groove was assumed to be sufficiently stiff 
(with very
high $K$). The main limitation of this model was the fact that it only worked
for high $K$, and therefore failed to produce any oscillations in a virtual
environment equivalent to water. At the time however, it was believed that the
worms swimming in water and crawling on agar were separate locomotory gaits
(like a horse's trot and gallop) \cite{white86}. 
One might therefore conjecture that the two behaviors would involve separate 
neural mechanisms. 

So how does an organism with such a limited neural circuit adapt its locomotion
waveform in environments with drastically different properties?  The fact that
the worm is capable of a continuous range of behaviors suggests that it relies
on a continuous modulation of a single mechanism grounded in a single circuit. 
The most parsimonious explanation would involve an effectively unchanged 
neural system, whose activity is modulated solely by the physics of the 
environment. This is not unlikely in a low Re environment.  In fact,
passive flagella of bacteria also exhibit shape changes when the viscosity
of their environment is increased \cite{schneider74}, and even more pronounced 
effects can be observed in the waveform of a sperm's flagellum (where active 
bending occurs along the length of the filament) \cite{brokaw66}.  
This suggests that a modulation of waveform by
physical forces is worth exploring. Within the context of a neural system,
one would expect a mechanism driven by sensory-feedback to be more naturally 
suited to direct modulation by the physics of the environment than a CPG 
controlled one. Furthermore, one would expect that without the constraints
of an internally clocked system, the modulation could take place over
a wider dynamic range.  Such a mechanism would need to explain
changes in frequency, amplitude and wavelength of undulations \cite{hfsp}.

It is not surprising that an increase in mechanical load on the body and
muscles (with increased viscosity or visco-elasticity) will result in slower
contraction. To explain the corresponding decrease in wavelength and amplitude,
it is convenient to imagine the body initiating an undulation from a straight
line configuration. The higher the environmental resistance, the harder it is
for the body to bend. This change is most pronounced in the
middle of the body, as bending here requires significant movement of the head
and tail. The head, which is strongly activated and has one free end, will
react first. At some point the head will be sufficiently bent that the stretch
receptors trigger and the direction of bending is reversed. The mid-body
meanwhile is still trying to bend in the first direction and will only reverse
after some delay.  (This assumes that during locomotion, a straight worm
is unstable and will tend to bend.) As the viscosity increases, the phase lag 
between adjacent parts of the body will increase, leading to a shorter and 
shorter wavelength.

Consider now the average body curvature which increases with viscosity or
visco-elasticity.  Assuming that stretch receptor signals are integrated over
some length of the body, the average curvature will be related to the
wavelength of the undulation \footnote{Indeed, worms moving in highly resistive
environments are observed to have higher average curvature, shorter wavelengths
and lower amplitudes than in water.}. Specifically, the higher the curvature,
the shorter the wavelength. When the wavelength is short, the stretch receptor
integrates over a significant fraction of the wavelength. Thus while some parts
of its receptive field will be highly curved, other parts will be nearly
straight, or even curved the other way. Conversely, when the wavelength is
long, the curvature will be more homogeneous along the receptive field.  Thus,
while the level of integrated curvature required to cause a reversal of bending
will be the same across the different environments, the peak curvature will be
greater when the wavelength is shorter.

The above reasoning was used in an integrated neuro-mechanical model of
nematode locomotion that reproduced the swim-crawl transition across different
environments \cite{boyle09}. This model is grounded in the known neural
circuitry (a variation on Fig.\ \ref{fig:cpgs}B), and does not contain a CPG
either in the head or along the body.  The only parameter changes required 
to modulate the locomotion
are to the drag coefficients $c_{\parallel}$ and $c_{\perp}$.  The key
ingredients of the model are bistability in the motor neurons (leading
to instability of straight line configurations) and sensory feedback via 
distributed stretch receptors. 

Having reviewed a system that can rely entirely on sensory (proprioceptive)
feedback to drive and modulate undulations, you might ask whether this esoteric 
example is of any interest to the neuroscience community, since, as mentioned 
above, such extreme reliance on sensory feedback and absence of CPGs is truly
unprecedented in the animal world. However, it is important to remember
that most living systems combine central control with sensory modulation
\cite{wilson61,pearson95,yu99}
and where advantageous, such systems will exploit the physics of the 
environment. Indeed, more and more, examples are being studied in which
sensory-motor control takes over when the CPG activity is somehow disabled
\cite{yu99}.

\section{Applications to mobile micro- and nanorobotics}

In this paper we have focused on a few examples of low Re undulatory
locomotion, but the principles, theories and techniques can all be applied to
many other situations. There are countless organisms that use undulatory
locomotion similar in form to that presented here, over a range of Reynolds
numbers and spanning some 6 orders of magnitude in size.  Moving to high
Reynolds number, for example, means that certain principles (like the scallop
theorem) no longer apply but others (like the requirement of symmetry breaking)
still do. Here we will begin by introducing some interesting biological
examples (taken from our own bodies) of systems that, while incapable of
locomotion {\it per se}, bear similarities to what we have covered so far. This
will be followed by some examples of biologically inspired robotic applications
of undulatory locomotion.

It is perhaps not surprising that biology has applied the principles of
undulatory locomotion to the task of moving substances around. One
example of this is peristalsis, the mechanism by which food is moved through
our digestive tract \cite{huizinga09}.  
Imagine taking a swimmer and attaching
it to a solid object at one end. What we are left with is essentially a pump.  
Our intestine, while anchored in place, generates waves of muscle contraction which 
pump semi-digested food through our system. The smooth muscle of our intestine can 
only contract inwards (reducing the radius), but by propagating coordinated waves 
of activity the food is propelled in one direction by the leading edge of the 
wave in much the same way that our hypothetical organism in Sec.\ \ref{sec:rigid} 
generates thrust.

Another type of biological pump can be found, among many places, in our trachea
(wind pipe). The inside of our trachea is lined with organelles called cilia,
which are genetically very closely related to eukaryotic flagella, and possess
the same 9+2 microtubule structure \cite{bray01}.  Compared to flagella, cilia
are generally shorter and are found in greater numbers. While a flagellum
generally produces thrust in the direction of its long axis (like a propeller),
cilia generally produce thrust perpendicular to their long axis (like oars).
Their asymmetric beating pattern \cite{marino82} is such that they minimize
drag on the forwards stroke and maximize drag on the backwards stroke, thereby
generating thrust along the surface from which they protrude. In so doing, they
move mucus and dirt away from our lungs ready for expulsion. In fact, cilia
perform other roles as well. In many cases they act as sensory organs
\cite{bray01} (such as the cilia in our ears), but that is outside the scope of
this paper.  Motile cilia are also found on some micro-organisms like protozoa
(unicellular eukaryotes), where they are responsible for locomotion. Their
behavior is similar to those in our lungs, except that the flow of fluid in one
direction propels the protozoa in the other.

One of the most exciting applications of the theory of undulatory propulsion in
recent years is to the field of bio-inspired robotics. Traditionally robots
have used wheels or tracks to get around, but neither of these are a great
solution to the problem of locomotion in heterogeneous and difficult terrain.
Given the success with which biological organisms move around, it is natural to
attempt to emulate these behaviors. In many cases, legged locomotion seems like
the ideal solution and recent advances suggest that legged robots may soon be
sufficiently advanced to handle real terrain. But there are also many cases
where undulatory locomotion would be preferable. A snake-like robot
\cite{hopkins09} may, for example, be able to inspect the inside of narrow
pipes, search for survivors among the rubble of collapsed buildings and
traverse even the roughest terrain. A snail-like robot (snail locomotion uses a
form of undulation) \cite{chan05} could climb vertically or even upside down.
But a particularly interesting environment in which small, worm-like robots
\cite{menciassi06} could potentially operate is the human body \cite{chen08}.
Designers of medical technologies constantly strive to find non-invasive (or
less invasive) ways to diagnose and repair problems within our bodies, and
micro-robots (or even remotely guided bacteria \cite{martel06}) are a promising
avenue. Undulatory locomotion \cite{behkam06} is ideally suited to use within
the human body as it is generally robust to heterogeneous environments and is
non-destructive.  While we are still a long way from this final goal,
researchers working on small scale undulatory robots are following a number
of interesting directions and making strides both on foundational issues 
and on the non-trivial engineering challenges of miniaturization.\\

\newpage
\section{Discussion}

In this paper, we have gone from reviewing some fundamental physics applied to
low Re swimmers, through different simulation approaches, to biological 
examples of undulating organisms. The field is too vast to have touched
on all the research that is relevant and ongoing, so we have chosen to
offer only a cursory view of better known systems and results (such
as bacterial swimming and lamprey CPGs), and to spend more time on
less familiar examples, such as Helmholtz's minimum energy dissipation
theorem, generalizations of the scallop theorem and nematode locomotion.

To the physicist uninitiated in biological parlance, the level of biological
detail in some of the examples may appear daunting, or perhaps superfluous. 
However, one of the key roles of the biological physicist is to tease out
the essential features of what can seem an almost arbitrarily complicated
system. Therefore, we have not shied away from leaving
the physics grounded in at least some of the biological systems in this 
review. We hope that this approach serves less as a hindrance and 
more as an invitation to the readers to engage in the larger endeavor
that is the interface between physics and biology.

\begin{acknowledgments} The authors are grateful for useful discussions
with Robert McNeill Alexander, Duncan Dowson, Sam Braunstein and 
Stefano Berri. NC acknowledges support from the EPSRC and BBSRC.
\end{acknowledgments}

\bibliography{ULbib}

% \clearpage
% \cleardoublepage
% \onecolumn

\end{document}